\documentclass[12pt]{iopart}
\usepackage{iopams}
\usepackage{graphicx}

\begin{document}

\title{Griffiths phase and spontaneous exchange bias in La$_{1.5}$Sr$_{0.5}$CoMn$_{0.5}$Fe$_{0.5}$O$_{6}$}

\author{A G Silva$^{1}$, K L Salcedo Rodr\'{i}guez$^{2,3}$, C P Contreras Medrano$^{3}$, G S G Louren\c{c}o$^1$, M Boldrin$^{1}$, E Baggio-Saitovitch$^{3}$, and L Bufai\c{c}al$^1$}

\address{$^1$ Instituto de F\'{i}sica, Universidade Federal de Goi\'{a}s, 74001-970 , Goi\^{a}nia, GO, Brazil}

\address{$^2$ Instituto de F\'{i}sica La Plata, Facultad de Ciencias Exactas, UNLP, CONICET, 1900, La Plata, Argentina}

\address{$^3$ Centro Brasileiro de Pesquisas F\'{\i}sicas, 22290-180, Rio de Janeiro, RJ, Brazil}

\ead{lbufaical@ufg.br}

\begin{abstract}

La$_{1.5}$Sr$_{0.5}$CoMn$_{0.5}$Fe$_{0.5}$O$_{6}$ (LSCMFO) compound was prepared by solid state reaction and its structural, electronic and magnetic properties were investigated. The material forms in rhombohedral $R\bar{3}c$ structure, and the presence of distinct magnetic interactions leads to the formation of a Griffiths phase above its FM transition temperature (150 K), possibly related to the nucleation of small short-ranged ferromagnetic clusters. At low temperatures, a spin glass-like phase emerges and the system exhibits both the conventional and the spontaneous exchange bias (EB) effects. These results resemble those reported for La$_{1.5}$Sr$_{0.5}$CoMnO$_{6}$ but are discrepant to those found when Fe partially substitutes Co in La$_{1.5}$Sr$_{0.5}$(Co$_{1-x}$Fe$_{x}$)MnO$_{6}$, for which the EB effect is observed in a much broader temperature range. The unidirectional anisotropy observed for LSCMFO is discussed and compared with those of resembling double-perovskite compounds, being plausibly explained in terms of its structural and electronic properties.
\end{abstract}

\section{Introduction}

The exchange bias (EB) effect is a remarkable phenomenon discovered at the 1950 decade in materials presenting ferromagnetic (FM)-antiferromagnetic (AFM) interfaces \cite{Bean}, and is manifested as a shift in the hysteresis loop measurement [$M(H)$]. For the recently discovered spontaneous EB (SEB) effect, the unidirectional anisotropy (UA) sets in spontaneously during the cycling of magnetic field ($H$) at the $M(H)$ measurement even when the material is cooled from an unmagnetized state down to low temperature ($T$) in zero $H$. This results from a field-induced symmetry breaking formed across different magnetic phases \cite{Wang,Nayak}. Conversely, for the conventional EB (CEB) effect discovered more than 60 years ago the asymmetry in the hysteresis loop is only manifested after the system is cooled in the presence of $H$ \cite{Nogues}. This phenomena is a key functional ingredient in the sensing elements of read heads, as well as in the magnetic media of hard disks. Therefore, the fact that no external cooling $H$ is needed to set the UA has raised to a great interest in the recently discovered SEB materials, where the presence of a spin glass (SG)-like phase seems to be a necessary condition for the appearance of SEB effect \cite{PRB2018,JMMM2020}. 

In this context, the double-perovskite (DP) compounds are very appealing since intrinsic structural and magnetic inhomogeneity usually lead to competing magnetic interactions and frustration, both key ingredients to achieve glassy magnetic behavior. This makes such materials promising candidates to exhibit SEB effect \cite{Sami,Mydosh}. For instance, the La$_{2-x}$A$_{x}$CoMnO$_{6}$ (A = Ba, Ca, Sr) DP family presents unequivocal SEB, with the $x$ = 0.5 Ba- and Sr-members showing a robust effect whilst for the Ca-one the shift in the $M(H)$ loop is subtle \cite{Murthy,Ba,JMMM2017}. Such differences are believed to be related to distinct amounts of low-spin (LS) Co$^{3+}$ present in each sample, which is strongly correlated to the structural particularities of the materials, leading to distinct uncompensation in the AFM coupling between Co and Mn \cite{PRB2019}.

Recently, it was shown that a Fe to Co partial substitution in La$_{1.5}$Sr$_{0.5}$CoMnO$_{6}$ yields a large SEB effect which changes sign depending on the concentrations of Fe and Co, such changes being ascribed to Co and Fe-dependent spin state transitions \cite{Zhang}. In this work we show that the Fe to Mn substitution also leads to unambiguous SEB effect. The structural, electronic and magnetic properties of La$_{1.5}$Sr$_{0.5}$CoMn$_{0.5}$Fe$_{0.5}$O$_{6}$ (LSCMFO) were thoroughly investigated by means of x-ray powder diffraction (XRD), M\"{o}ssbauer spectroscopy and magnetization as a function of $H$ and $T$ measurements, evidencing the presence of FM and AFM phases. The magnetic inhomogeneity leads to a Griffiths-like phase at high $T$ and to a SG-like phase at low $T$. Both SEB and CEB effects are observed below 10 K in this material. The magnetic properties of LSCMFO are compared to those of similar DP compounds and explained in terms of its structural and electronic properties.

\section{Experiment details}

A polycrystalline sample of LSCMFO was synthesized by conventional solid-state reaction method. Stoichiometric amounts of La$_{2}$O$_{3}$, SrCO$_3$, Co$_{3}$O$_{4}$, MnO and Fe$_{2}$O$_{3}$ were mixed and heated at $750^{\circ}$C for 10 hours in air atmosphere. Later the sample was re-grinded before a second step at $1400^{\circ}$C for 36 hours, with intermediate grinding. Finally, the material was grinded, pressed into pellet and heated at $1400^{\circ}$C for 24 hours. After this procedure it was obtained a dark-black material in the form of a 10 mm diameter disk. High resolution XRD data was collected at room $T$ using a Bruker $D$8 $Discover$ diffractometer, operating with Cu $K_{\alpha}$ radiation operating at 40 kV and 40 mA. The XRD data was investigated over the angular range $10\leq\theta\leq90^{\circ}$, with a 2${\theta}$ step size of 0.01$^{\circ}$. Rietveld refinement was performed with GSAS software and its graphical interface program \cite{GSAS}. $^{57}$Fe-M\"{o}ssbauer measurement was performed in transmission geometry at room temperature (RT), with a $^{57}$Co:Rh source moving in a sinusoidal motion. Experimental data were fitted with the Normos program. 

Magnetization as a function of $T$ [$M(T)$] and $M(H)$ measurements were carried out in both zero field cooled (ZFC) and field cooled (FC) modes using a commercial SQUID-VSM magnetometer. 
The $M(H)$ measurements were performed at several $T$ and, from one experiment to another, a systematic protocol of sending the sample to the paramagnetic (PM) state and demagnetize the system in the oscillating mode was adopted in order to prevent the presence of remanent magnetization. This procedure is particularly important for the ZFC $M(H)$ measurements, since any trapped current in the magnet would prevent the correct observation of the SEB effect.

\section{Results and discussion}

\begin{figure}
\begin{center}
\includegraphics[width=0.5 \textwidth]{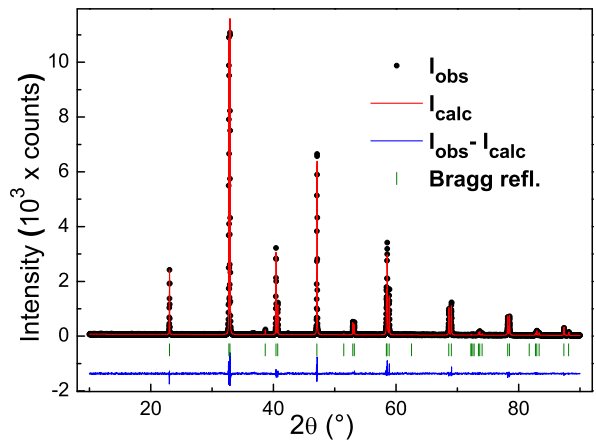}
\end{center}
\caption{Rietveld refinement fitting of LSCMFO. The vertical lines represent the Bragg reflections for the $R\bar{3}c$ space group.}
\label{Fig_XRD}
\end{figure}

The room-$T$ XRD pattern of LSCMFO is shown in Fig. \ref{Fig_XRD}. It revealed the formation of a single phase perovskite compound and could be successfully refined in the rhombohedral $R\bar{3}c$ space group, as for La$_{1.5}$Sr$_{0.5}$(Co$_{1-x}$Fe$_x$)MnO$_{6}$ and other resembling compounds\cite{Murthy,PRB2019,Zhang}. The main parameters obtained from the Rietveld refinement are displayed in Table \ref{T1}. 

Comparing the lattice parameters of LSCMFO with those reported for La$_{1.5}$Sr$_{0.5}$CoMnO$_{6}$ (LSCMO) \cite{Murthy2} it is observed a subtle increase in the unit cell volume of the former (344.09 \AA$^{3}$) in relation to the later compound (343.58 \AA$^{3}$). This was expected since our M\"{o}ssbauer measurements presented below indicate trivalent state for Fe ions and previous works have shown a nearly tetravalent state for the Mn ions in LSCMO \cite{Murthy,PRB2019,Murthy2}. Since the Mn$^{4+}$ ionic radius is smaller than that of Fe$^{3+}$ \cite{Shannon}, the incorporation of Fe in the system is expected to expand the lattice, as observed. On the other hand, previous studies have also indicated a Co$^{2+}$/Co$^{3+}$ mixed valence state in LSCMO and resembling compounds \cite{PRB2019,Murthy2,Arkadeb2}. Therefore, the introduction of Fe$^{3+}$ in place of Mn$^{4+}$ induced the Co ions to a trivalent state in order to ensure the charge balance, which prevents a larger increase of the unit cell volume. For this scenario, we are assuming a near stoichiometry of oxygen ions, which may not be precisely the case since excess or vacancy of oxygen ions usually occurs in perovskites and should be verified by means of thermogravimetric analysis, X-ray photoelectron spectroscopy or X-ray absorption spectroscopy. Nevertheless, the Co, Fe and Mn formal valences here described are in agreement with the magnetization results.

The B--O bond length (B = Co/Mn/Fe) is slightly smaller for LSCMFO (1.941 \AA) than for LSCMO (1.946 \AA) \cite{Murthy2}, which may be related to a more pronounced tilt of the oxygen octahedra for the later compound. Although care must be taken in interpreting these data since they depend on the oxygen positions taken from XRD, the trends here described are highly expected for perovskite compounds \cite{Sami} and are supported by our magnetization data, being thus very plausible. The structural and electronic changes observed for LSCMFO with respect to LSCMO will remarkably affect the compound magnetic properties, as will be discussed below.

\begin{table}
\caption{Structural and magnetic parameters obtained from XRD and $M(T)$ data, and M\"{o}ssbauer parameters of the paramagnetic subspectra $D_1$ and $D_2$ \label{T1}}
\begin{tabular}{cccccc}
\hline
\hline
$a$ (\AA)  & $c$ (\AA)  & V (\AA$^{3}$) & B--O (\AA) & $\chi^{2}$  & $R_{wp}$   \\
 5.4682(1) & 13.2879(1) & 344.09(1) & 1.941(1) & 1.7 & 13.1 \\
\hline 
 $T_C$ (K) & $T_{g}$ (K) & $\theta_{CW}$ (K) & $\mu_{eff}$  ($\mu_{B}$/f.u.) & $\lambda$ & $T_G$ (K) \\
 150 & 73 & 59 & 5.2 & 0.64 & 205 \\
\hline
   & $\delta$ (mm/s)  & $\Delta$E$_Q$ (mm/s) & $\Gamma$ (mm/s) & A (\%) &  \\
$D_1$ & 0.33 & 0.67 & 0.31 & 40  & \\
$D_2$ & 0.34 & 0.34 & 0.30 & 60  & \\ 
\hline
\hline
\end{tabular}
\end{table}

\begin{figure}
\begin{center}
\includegraphics[width=0.5 \textwidth]{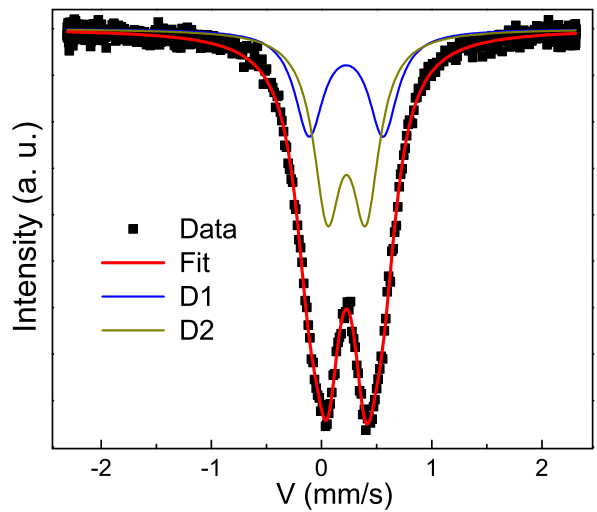}
\end{center}
\caption{$^{57}$Fe M\"{o}ssbauer spectrum at RT showing two different Fe$^{3+}$ sites.}
\label{Fig_Mossbauer}
\end{figure}

$^{57}$Fe M\"{o}ssbauer spectrum is displayed in Fig. \ref{Fig_Mossbauer}. To fit this spectrum, a model with two paramagnetic doublets gives the best least-square fitting. The obtained hyperfine parameters are the isomer shift ($\delta$), the quadrupole splitting ($\Delta$E$_Q$), the spectral relative area (A) and the line width ($\Gamma$). Table \ref{T1} presents the hyperfine parameters, $\delta$ values are reported relative to the $\alpha$-Fe. For these subspectra, the values of $\delta$ and $\Delta$E$_Q$ are consistent with Fe$^{3+}$ at octahedral sites. The electric field gradient produced by the surrounding atoms at iron probe sites gives rise to the quadrupole splitting \cite{Greenwood,McCammon}. In a perfect cubic symmetry, $\Delta$E$_Q$ = 0, other values indicate the deviation of the system from a cubic symmetry. Despite a signature of disorder is showed in the linewidth, the subspectra of two doublets have values of $\delta$ and $\Delta$E$_Q$ consistent with Fe$^{3+}$ at octahedral sites. Then, the presence of two paramagnetic components could be related to the statistical distribution of the atomic environment around Fe$^{3+}$, being the main part more symmetric. This was expected since in this disordered system one may have regions with Fe surrounded by Co and other ones where Mn is its nearest neighbor. More information could be obtained from the M\"{o}ssbauer spectra below the magnetic order as observed for other perovskite compounds \cite{Stoch,Gibb}

\begin{figure}
\begin{center}
\includegraphics[width=0.5 \textwidth]{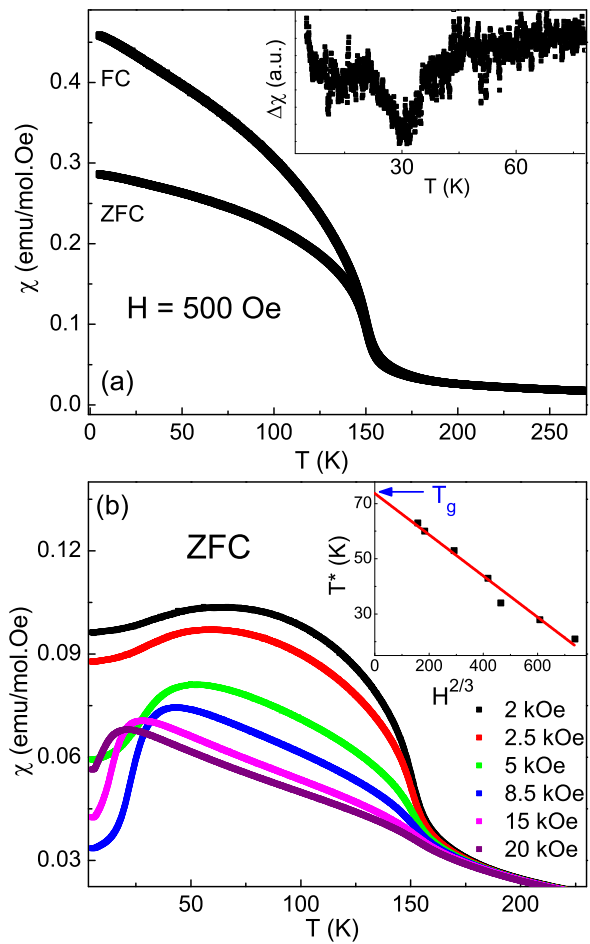}
\end{center}
\caption{(a) ZFC-FC $\chi(T)$ at $H$ = 500 Oe. The inset shows the memory effect for the $H$ = 500 Oe $\chi(T)$ ZFC curve halted at 30 K (see text). (b) Magnified view of the low-$T$ region of ZFC $M(T)$ curves measured with different $H$. The inset shows the $T$ of maxima in the curves as a function of $H^{2/3}$, and the solid red line represents the best linear fit.}
\label{Fig_MxT}
\end{figure}

Fig. \ref{Fig_MxT}(a) shows the ZFC and FC $M(T)$ curves measured with $H$ = 500 Oe, from which is observed a FM-like transition at $T_{C}\simeq$150 K. The fit of the inverse of magnetic susceptibility ($\chi^{-1}$) with the Curie-Weiss (CW) law at the PM region yielded a CW temperature $\theta_{CW}$ = 59 K and an effective magnetic moment $\mu_{eff}$ = 5.2 $\mu_B$/f.u.. The positive $\theta_{CW}$ confirms the presence of FM coupling. However, the fact that $\theta_{CW}$ lies far below $T_C$, together with the small magnetization values observed at low-$T$, rule out the possibility of a fully long-ranged FM state with the presence of Fe, Co and Mn magnetic ions. According to the Goodenough-Kanamory-Anderson (GKA) rules \cite{Goodenough}, the Fe$^{3+}$--O--Mn$^{4+}$ coupling is expected to be FM, which could explain the positive $\theta_{CW}$ and the FM-like curve depicted in Fig. \ref{Fig_MxT}(a). On the other hand, for the $R\bar{3}c$ space group all Co, Fe and Mn ions share the same crystallographic site, which is intrinsically related to the proximity of Co/Mn/Fe ionic radii and scattering factors. This results in a complex and disordered system with a high degree of permutation between the transition-metal (TM) ions, for which other nearest neighbor couplings as Fe$^{3+}$--O--Fe$^{3+}$, Mn$^{4+}$--O--Mn$^{4+}$ are expected. These are predicted by the GKA rules to be of AFM type, and in such condition the superexchange-driven FM contribution is interrupted by a large proportion of AFM phases, which explain the small magnetization observed for LSCMFO.

As aforementioned, the Fe$^{3+}$ to Mn$^{4+}$ partial substitution drives the Co toward a trivalent state. This ion is particularly interesting due its multiple spin states resulting from a delicate balance between between the crystal field splitting and the intra-atomic Hund’s exchange energy \cite{Raveau}. In octahedral coordination, Co$^{3+}$ can be found in high spin (HS) \cite{Shukla,Kumar}, low spin (LS) \cite{PRB2019,Zhang,Arkadeb,Baier} or even in intermediate spin (IS) configuration \cite{Fauth}. As a first attempt to reproduce the experimentally observed $\mu_{eff}$, one can consider the HS configuration for Co$^{3+}$. Using the spin only approximation for the TM ions, \textit{i.e.} $\mu_{Fe^{3+}}$ = 5.9 $\mu_B$, $\mu_{Mn^{4+}}$ = 3.9 $\mu_B$ and $\mu_{HSCo^{3+}}$ = 4.9 $\mu_B$, the effective magnetic moment can be theoretically calculated with the following equation for systems with two or more different magnetic ions \cite{Niebieskikwiat,JSSC2015}
\begin{equation}
\mu = \sqrt{{\mu_{HSCo^{3+}}}^2 + 0.5{\mu_{Fe^{3+}}}^2 + 0.5{\mu_{Mn^{4+}}}^2}, \
\end{equation}
yielding $\mu$ = 7.0 $\mu_B$/f.u., far above the experimentally observed value. Now assuming LS Co$^{3+}$ (S = 0) while keeping Mn$^{4+}$ and Fe$^{3+}$, the calculated magnetic moment becomes $\mu$ = 5.0 $\mu_B$/f.u., much closer to the experiment.

H. G. Zhang \textit{et al.} reported the presence of a second FM transition for intermediate Co/Fe concentrations in the La$_{1.5}$Sr$_{0.5}$(Co$_{1-x}$Fe$_{x}$)MnO$_6$ series \cite{Zhang}, finding that was not confirmed in the $M(T)$ curves of the La$_{1.5}$Sr$_{0.5}$Co(Fe$_{0.5}$Mn$_{0.5}$)O$_6$ sample here investigated. Since both systems belong to $R\bar{3}c$ space group, for which the Co/Fe/Mn ions share the same crystallographic site, this different magnetic behavior is most likely related to changes in valence and spin states, which in turn are related to the molar ratios among these ions. The Co:Fe:Mn ratio for the latter system is (1-x):x:1, leading to mixed valence states for Mn (Mn$^{3+}$/Mn$^{4+}$) while Co presents both mixed valence and spin states (Co$^{2+}$/Co$^{3+}$ in LS and HS configurations) for intermediate $x$. The second transition was attributed to FM exchange couplings as (HS)Co$^{3+}$--O--Mn$^{4+}$, (LS)Co$^{2+}$--O--(HS)Co$^{3+}$, (LS)Co$^{2+}$--O--Fe$^{3+}$ and (HS)Co$^{3+}$--O--(LS)Co$^{2+}$ \cite{Zhang}. Contrastingly, the Co:Fe:Mn ratio for La$_{1.5}$Sr$_{0.5}$Co(Fe$_{0.5}$Mn$_{0.5}$)O$_6$ is 1:0.5:0.5, with this compound being better described as a single valence and single spin state system [Mn$^{4+}$, (LS)Co$^{3+}$], and naturally the second FM transition is not observed.

Fig. \ref{Fig_MxT}(a) shows a split of the ZFC and the FC $M(T)$ curves below 150 K. This is characteristic of disordered magnetic systems, where the presence of competing magnetic phases usually leads to glassy magnetic behavior \cite{Mydosh}. Due to this, an interesting anomaly is observed in the $M(T)$ curves carried with $H$ $\geq$ 2000 Oe, Fig. \ref{Fig_MxT}(b). A peak appears in the low-$T$ region, whose intensity decreases with increasing $H$ while it moves toward lower $T$. These are characteristic features of SG-like states, here resulting from the structural disorder together with competition between FM and AFM phases. The fact that the anomaly doesn't manifest for small $H$ suggests that larger fields are necessary for one of the competing magnetic interactions to overmatch the other, thus indicating that the cluster glass (CG) state is most probable in LSCMFO \cite{Mukadam,Mukherjee,Freitas,Kleemann}. The inset of Fig. \ref{Fig_MxT}(b) shows the evolution of the peak position ($T^{*}$) as a function of the measured $H$. The $T^{*}$, obtained from the first derivative of the ZFC $M(T)$ curves, reasonably follows the $H^{2/3}$ Almeida-Thouless (AT) relation usually observed for SG-like systems \cite{AT}, further indication that this peak may be associated to the onset of glassy magnetic behavior. From the extrapolation of the linear fit shown in the inset, the freezing $T$ of the possibly CG phase in zero $H$ would be $T_g$ $\simeq$ 73 K. The formation of the clusters, as well as their sizes, should be confirmed by other techniques as AC magnetic susceptibility, neutron powder diffraction and electron spin resonance.

\begin{figure}
\begin{center}
\includegraphics[width=0.5 \textwidth]{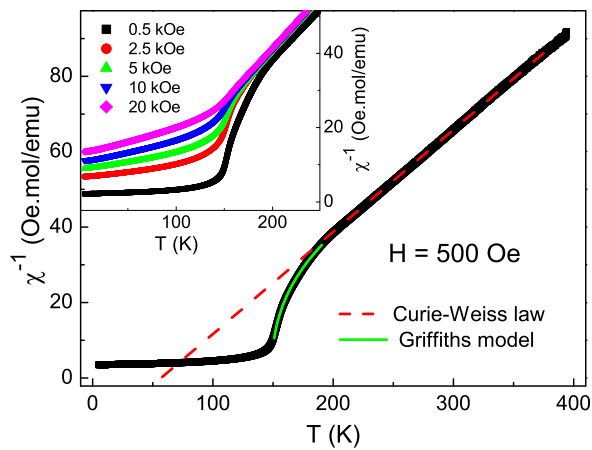}
\end{center}
\caption{$\chi^{-1}$ as a function of $T$ at $H$ = 500 Oe. The red dashed line is the fit with the CW law and the solid green line is the fit with the Griffiths model. The inset shows $\chi^{-1}$ at different $H$.}
\label{Fig_GP}
\end{figure}

In order to further investigate the SG-like state in LSCMFO, and to confirm that the glassy behavior is present even when the anomaly is not apparent in the $M(T)$ curves (\textit{i.e.} for $H<2000$ Oe), we have performed a memory effect measurement in the $H$ = 500 Oe ZFC susceptibility by adopting the following protocol: first, a standard ZFC curve was captured by cooling the system down to 5 K in zero field for a subsequent application of $H$ = 500 Oe and the measurement of $\chi$$vs$$T$ in the heating mode. In the second step a similar measurement was performed, but with the difference that during the ZFC protocol the cooling procedure was paused for a time-interval $t_w$ = 10$^{4}$ s at $T$ = 30 K. The inset of Fig. \ref{Fig_MxT}(a) shows the difference between these curves, where a dip is observed precisely at 30 K. This is a hallmark of glassy magnetic systems, thus confirming the emergence of SG-like state at low-$T$.

Fig. \ref{Fig_GP} shows $\chi^{-1}$ of the $H$ = 500 Oe ZFC curve. One can note a rapid downward deviation from the CW law occurring well-above $T_C$, which is a signature of Griffiths phase (GP). This phenomena is related to the nucleation of small and weakly correlated clusters embedded in a PM matrix, resulting in a peculiar magnetic phase where the system neither behave like a PM nor shows long-range ordering \cite{Griffiths,Bray}. The inset of Fig. \ref{Fig_GP} shows that as $H$ increases the downturn deviation gets softened. This is also a characteristic feature of GP, related to the fact the PM matrix magnetization increases linearly with increasing $H$ \cite{Salamon,Palakkal}, and for a sufficiently high field its susceptibility dominates over the weakly correlated clusters. This scenario of short-ranged FM clusters may explain the small magnetization values observed in the $M(T)$ and $M(H)$ curves.

The characteristics of the GP are usually described by the following power law equation \cite{Salamon}
\begin{equation}
\chi^{-1} \propto (T - T_{C}^{R})^{1-\lambda}, \
\end{equation}
which is simply a modified version of the CW law where the value of $\lambda$ (0$\leq\lambda\leq$1) reflects the strength of the GP, \textit{i.e.} it is a measure of the deviation from the CW behavior. $T_{C}^{R}$ defines the GP interval, $T_{C}^{R}<T<T_G$, with $T_G$ being the temperature above which the system behaves as a simple PM. The $\lambda$ value is highly sensitive to $T_{C}^{R}$ \cite{Arkadeb,Pramanik}, which must be carefully computed in order to obtain a reliable description of the GP. In the case of LSCMFO, since the $\theta_{CW}$ obtained from the CW fitting of the PM region lies far below the FM ordering $T$, we have chosen $T_{C}^{R}=T_C$, yielding $\lambda$ = 0.64. This value lies within the range found for resembling DP compounds \cite{Zhang,Arkadeb2,Arkadeb,Elizabeth}. The lattice distortions and the cationic disorder at the TM ion-site, intrinsic to the $R\bar{3}c$ space group,  certainly play a part in the GP. But the abundant non-magnetic LS Co$^{3+}$ is probably the key for the appearance of such phenomena in LSCMFO as it gives rise to the dilution of the magnetic ions, thus hindering the formation of long-range order and aiding the nucleation of correlated clusters. 

\begin{figure}
\begin{center}
\includegraphics[width=0.5 \textwidth]{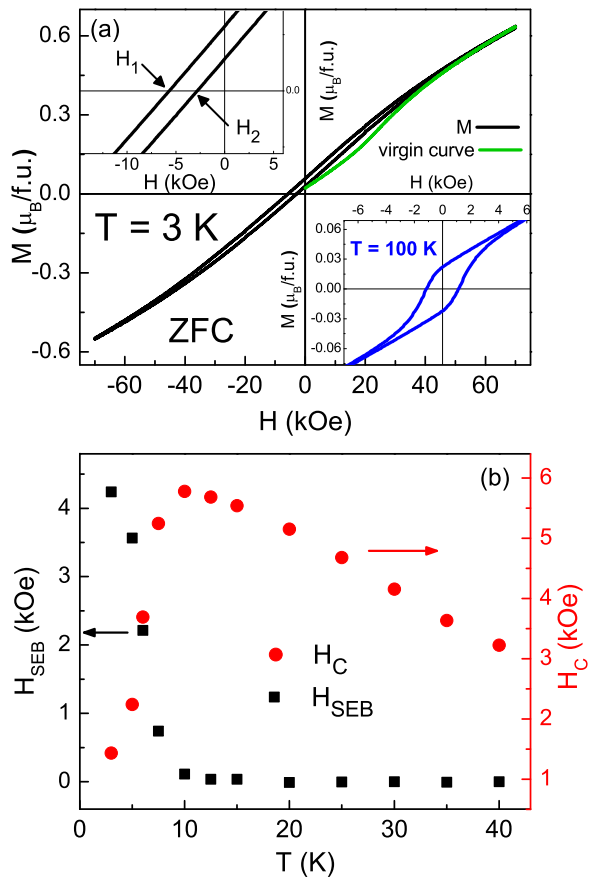}
\end{center}
\caption{(a) ZFC $M(H)$ loop at 3 K, measured with $H_{max}$ = 70 kOe. The upper inset shows a magnified view of the regions close to the coercive fields, highlighting the SEB effect. The bottom inset shows the $M(H)$ loop at 100 K. (b) $H_{SEB}$ and $H_C$ evolution with $T$.}
\label{Fig_SEB}
\end{figure}

It is well known that multi-magnetic systems usually exhibit the EB effect \cite{Nogues}, and recently it was shown that glassy magnetism is paramount for the emergence of SEB \cite{PRB2018,JMMM2020}. Thereby such phenomenon was verified in LSCMFO. Fig. \ref{Fig_SEB}(a) shows the $M(H)$ curve measured after ZFC the sample down to 3 K. It is clearly a closed loop, nearly symmetric with respect to the $M$ axis and shifted to the left along the $H$ axis. It may also be noticed that the virgin magnetization curve falls outside the hysteresis loop up to a certain critical $H$ ($\sim$4.5T), which is a common ground of several SEB compounds, being related to $H$ induced magnetic transitions  \cite{Wang,Nayak,Murthy,JMMM2017}. Since the 3 K $M(H)$ curve of LSCMFO results from the contributions of short-ranged FM clusters embedded in a AFM matrix, there is no magnetic saturation up to 70 kOe. Furthermore, by comparing this curve with the one taken at 100 K [bottom inset of Fig. \ref{Fig_SEB}(a)] one can see that for the lower $T$-curve the loop loses it squareness, indicating that the system has entered in the glassy state.

The shift along the $H$ axis is a measure of the EB field, herein defined as $H_{EB}=|H_{2}+H_{1}|/2$, where $H_1$ and $H_2$ are respectively the left and right coercive fields (keeping the signs). The average coercive field is defined as $H_C=(H_{2}-H_{1})/2$. For the 3 K curve shown in Fig. \ref{Fig_SEB}(a) it was found a maximum shift of 4241 Oe, which is comparable to those of giant SEB materials \cite{Murthy,Nath}. Several ZFC $M(H)$ measurements were performed at different $T$ in order to investigate the evolution of the SEB effect. It is shown in Fig. \ref{Fig_SEB}(b) that the SEB effect is negligible for $T>10$ K, the drop in $H_{SEB}$ being accompanied by a local increase of $H_C$. This is a signature of the EB effect \cite{Nogues},  where the gain in thermal energy favours some AFM spins to be ``dragged" by the rotating FM clusters, thus increasing $H_C$. 

\begin{figure}
\begin{center}
\includegraphics[width=0.5 \textwidth]{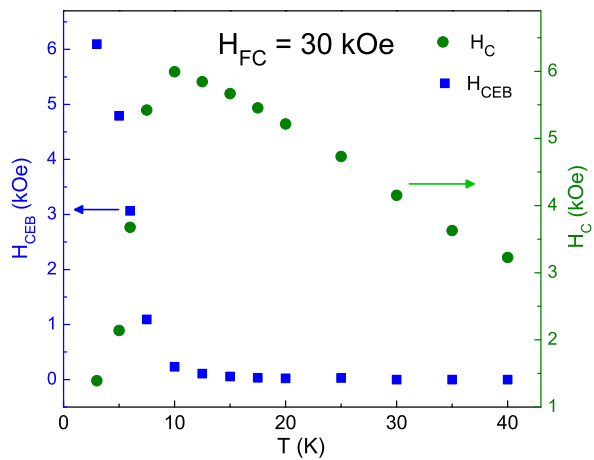}
\end{center}
\caption{Evolution of $H_{CEB}$ and $H_C$ with $T$ for $M(H)$ curves measured with $H_{max}$ = 70 kOe after cooling the sample in the field $H_{FC}$ = 30 kOe.}
\label{Fig_CEB}
\end{figure}

There is a great increase in the EB effect of LSCMFO when the system is cooled in the presence of an external field. For a cooling field $H_{FC}$ = 30 kOe the shift in the 3 K $M(H)$ curve is $\sim$40\% larger than that of the ZFC loop measured at the same $T$, as can be seen in Fig. \ref{Fig_CEB}. The qualitative evolution of both $H_{CEB}$ and $H_C$ as a function of $T$ are very similar to those observed after the ZFC procedure. Once again, the emergence of EB at low $T$ indicates that the glassy magnetic phase plays a key role on the EB effect of DP compounds. The overall CEB behavior of LSCMFO resembles the results observed for LSCMO and related compounds, for which $H_{FC}$ induces the growth and the correlation between the FM clusters, favoring the pinning of spins and thus increasing the EB effect \cite{Murthy,JMMM2017,PRB2016}.

\begin{figure}
\begin{center}
\includegraphics[width=0.5 \textwidth]{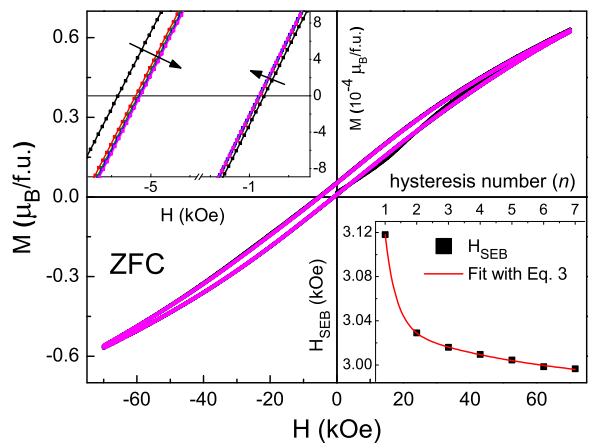}
\end{center}
\caption{Seven successive $M(H)$ cycles, measured at 5 K after ZFC the sample. The upper inset shows a magnified view of the regions close to $M$ = 0, highlighting the changes in $H_1$ and $H_2$. The bottom inset shows $H_{SEB}$ as a function of the hysteresis number ($n$).}
\label{Fig_TE}
\end{figure}

In EB systems, $H_{EB}$ generally depends on the number $n$ of consecutive $M(H)$ measurements, a property often called training effect \cite{Nogues}. This is usually related to the relaxation of uncompensated spins after successive $M(H)$ cycles, and for practical applications in spintronics it is required of a EB material to be fairly stable under successive $H$ cycles. Here we investigated the changes in the SEB effect for seven consecutive hysteresis loops. Fig. \ref{Fig_TE} shows that the consecutive cycles lead to very small changes, evidencing the stability of the SEB effect. A closer inspection of the curves near the $M=0$ region (upper inset of Fig. \ref{Fig_TE}) indicates a monotonic decrease of the left coercive field $H_1$ with increasing the number of cycles $n$, suggesting that some spin rearrangements at the magnetic interfaces lead to the reduction of $H_{SEB}$. The decrease is more pronounced from the first to the second cycle, resulting in a larger reduction of the EB effect from $n$ = 1 to $n$ = 2. Conversely, the right coercive field $H_2$ systematically shifts to the left with increasing $n$. This acts in favor of increasing the EB effect, but since the changes are more pronounced in $H_1$ than in $H_2$ the consecutive $M(H)$ loops act to slowly reduce $H_{SEB}$ with increasing $n$. The decrement from $n$ = 1 to $n$ = 2 is $\sim$3\% and to $n$ = 7 is  $\sim$4\%, these very small values indicating that the metastable spin configuration at the magnetic interfaces is reasonably stable against the magnetic field cycles.

The $H_{SEB}$ evolution with $n$ could be well fitted by a model proposed for FM/SG systems which considers that the SG state has both frozen and uncompensated rotatable spins \cite{PRB2016,Mishra}
\begin{equation}
{H}_{SEB}(n) = H^{\infty}_{SEB} + A_{f}e^{(-n/P_{f})} + A_{r}e^{(-n/P_{r})}, \label{EqTE}
\end{equation}
where $H^{\infty}_{SEB}$ is the value of $H_{SEB}$ for $n$ = $\infty$, $A_{f}$ and $P_f$ are parameters related to the change of the frozen spins, $A_{r}$ and $P_r$ are evolving parameters of the rotatable spins. The fitting with  Eq. \ref{EqTE} yields $H^{\infty}_{SEB}$ $\simeq$ 2981 Oe, $A_{f}$ $\simeq$ 1422 Oe, $P_{f}$ $\simeq$ 0.35, $A_{r}$ $\simeq$ 66 Oe, $P_{r}$ $\simeq$ 4.82. The precise fitting of the training effect with Eq. \ref{EqTE} gives further evidence of the presence of SG-like phase in LSCMFO. As observed for other DPs, the glassy magnetism seems to play a key role on the SEB of LSCMFO.

\section{Summary}

In summary, the structural, electronic and magnetic properties of LSCMFO compound were systematically studied. The polycrystalline sample forms in rhombohedral $R\bar{3}c$ space group, as observed for similar compounds. The M\"{o}ssbauer spectroscopy has revealed that Fe is in trivalent oxidation state. The $M(T)$ data indicate a FM transition at 150 K related to the Fe$^{3+}$--O--Mn$^{4+}$ coupling, whilst several other possible nearest neighbor interactions as Fe$^{3+}$--O--Fe$^{3+}$ and Mn$^{4+}$--O--Mn$^{4+}$ are predicted to be AFM. Magnetization results also indicate that the presence of non-magnetic LS Co$^{3+}$ prevents the formation of long-range order, aiding the nucleation of FM clusters that, when embedded in a PM matrix at high $T$, drives the system to exhibit a GP. At low $T$, the onset of AFM couplings leads to the emergence of a CG state. The material exhibit a large SEB effect below 10 K, which is greatly enhanced when the system is cooled in the presence of an applied magnetic field. 

\section*{Acknowledgments}
This work was supported by Conselho Nacional de Desenvolvimento Cient\'{i}fico e Tecnol\'{o}gico (CNPq) [No. 400134/2016-0, 425936/2016-3  and 309250/2011-0], Coordena\c{c}\~{a}o de Aperfei\c{c}oamento de Pessoal de N\'{i}vel Superior (CAPES)and Funda\c{c}\~{a}o de Amparo \`{a} Pesquisa do Estado de Goi\'{a}s (FAPEG). E.B.S. aknowledges several grants from Funda\c{c}\~{a}o Carlos Chagas Filho de Amparo \`{a} Pesquisa do Estado do Rio de Janeiro (FAPERJ), including Professor Emeritus fellowship. C.P.C.M.  acknowledges FAPERJ for the PDR Nota10 fellowship and  K.L.S.R. thanks CNPq for PDJ fellowship.

\end{document}